\documentclass[prx,twocolumn,superscriptaddress]{revtex4-2}

\usepackage{upgreek}   %  font packages  字体
\usepackage{amsmath,amsthm,mathtools}
\usepackage{amssymb,amsfonts}
\usepackage{graphicx,floatrow, float}   %  figure packages  图片
\usepackage{multirow}   %  table packages  表格
\usepackage{}   %  format packages  格式
\usepackage{comment}   %  other packages  其他

\usepackage[colorlinks=true,linkcolor=black,citecolor=blue,urlcolor=blue,]{hyperref}
\usepackage[doipre={doi:~}]{uri}

\newcommand{\ket}[1]{|#1\rangle}

\begin{document}

\title{Observation of hierarchy of Hilbert space ergodicities in the quantum dynamics of a single spin}

\author{Wenquan Liu}\thanks{These authors contributed equally to this work.}
\affiliation{Institute of Quantum Sensing, Institute of Fundamental and Transdisciplinary Research, School of Physics, and State Key Laboratory of Ocean Sensing, Zhejiang University, Hangzhou, 310058, China}
\author{Zouwei Pan}\thanks{These authors contributed equally to this work.}
\affiliation{Institute of Quantum Sensing, Institute of Fundamental and Transdisciplinary Research, School of Physics, and State Key Laboratory of Ocean Sensing, Zhejiang University, Hangzhou, 310058, China}
\author{Yue Fu}
\affiliation{School of Integrated Circuits and Electronics, Beijing Institute of Technology, Beijing 100081, China}
\author{Wei Cheng}
\affiliation{Quantum Science Center of Guangdong-Hong Kong-Macao Greater Bay Area (Guangdong), Shenzhen 518045, China}
\author{Wen Wei Ho}
\email{wenweiho@nus.edu.sg}
\affiliation{Department of Physics, National University of Singapore, Singapore 117551, Singapore}
\affiliation{Centre for Quantum Technologies, National University of Singapore, 3 Science Drive 2, Singapore 117543, Singapore}
\author{Xing Rong}
\email{xrong@ustc.edu.cn}
\affiliation{Institute of Quantum Sensing, Institute of Fundamental and Transdisciplinary Research, School of Physics, and State Key Laboratory of Ocean Sensing, Zhejiang University, Hangzhou, 310058, China}
\affiliation{Laboratory of Spin Magnetic Resonance, School of Physical Sciences, Anhui Province Key Laboratory of Scientific Instrument Development and Application, University of Science and Technology of China, Hefei, 230026, China}
\affiliation{Hefei National Laboratory, University of Science and Technology of China, Hefei, 230088, China}
\author{Jiangfeng Du}
\affiliation{Institute of Quantum Sensing, Institute of Fundamental and Transdisciplinary Research, School of Physics, and State Key Laboratory of Ocean Sensing, Zhejiang University, Hangzhou, 310058, China}
\affiliation{Laboratory of Spin Magnetic Resonance, School of Physical Sciences, Anhui Province Key Laboratory of Scientific Instrument Development and Application, University of Science and Technology of China, Hefei, 230026, China}
\affiliation{Hefei National Laboratory, University of Science and Technology of China, Hefei, 230088, China}

\begin{abstract}
Ergodicity, the property that all allowed configurations are explored over time, plays a pivotal role in explaining the equilibrium behavior of classical dynamical systems. Yet, such a property is typically precluded in quantum systems owing to the presence of energy eigenstates, which are stationary states in dynamics. However, recent theoretical works have argued that ergodic explorations of the Hilbert space, occurring at varying levels as measured by statistical pseudorandomness of the time-evolved quantum states, may be exhibited for quantum systems driven by Hamiltonians with aperiodic time dependencies, which do not face such obstacles. Here, we experimentally investigate the hierarchy of Hilbert-space ergodicities (HSE) achievable in the dynamics of a single quantum spin realized by a solid-state defect in diamond, upon subjecting it to various time-dependent modulations. Through continuous monitoring of spin trajectories with full state tomography, different degrees of HSE were observed, ranging from no HSE in a time-periodic (Floquet) drive, to partial HSE in a smoothly kicked time-quasiperiodic drive, to complete HSE in a drive composed of a sequence of kicks generated by the Fibonacci word. 
We formulate a theoretical understanding of the increasing levels of HSE observed by attributing them to increasing levels of complexities associated with the drive sequences, whose notions we elucidate. 
Our work constitutes the first unambiguous experimental evidence of Hilbert space ergodicity and promotes deeper investigations into the mechanisms and fine-grained levels with which closed quantum systems reach equilibrium.
\end{abstract}
\maketitle

%%%%% introduction %%%%%%%%%%%%%%%%%%%%%%%%%%%%%%%%%%%%
\section{Introduction}
Ergodicity in classical systems refers to the dynamical property of the system exploring its allowed phase space uniformly over time (with respect to a natural invariant measure), regardless of the initial state~\cite{Boltzmann1910, Ollagnier1985, Arnold1989}. It forms the cornerstone of statistical physics, explaining how equilibrium arises from complex equations of motion~\cite{Callen1985, Sethna2012}. In the quantum realm, on the other hand, the idea of ergodicity is formulated rather differently~\cite{Rigol2008, Eisert2015, Nandkishore2015, Neill2016, Gogolin2016, Mori2018}. Owing to the fact that energy eigenstates of a quantum Hamiltonian are stationary states in dynamics, an initial-state-independent dynamical notion of ergodicity is not possible, and instead, quantum ergodicity is typically formulated in terms of the energy eigenstates' statistical similarity to random vectors~\cite{Luitz2016, Luitz2017, Wilming2019, Serbyn2021, Srdinsek2024}. Indeed, examples include Berry's conjecture in single-particle chaotic systems~\cite{Berry1977, Jarzynski1997}, and the eigenstate thermalization hypothesis in many-body interacting systems~\cite{Deutsch1991, Srednicki1994, Steinigeweg2013, Ikeda2013, Kim2014, Reimann2015, D’Alessio2016, Deutsch2018, Dymarsky2018}. 

Recent theoretical works, however, have argued that such a framework is not exhaustive: a dynamical formulation of quantum ergodicity may be recovered in quantum systems driven by Hamiltonians $H(t)$ with {\it aperiodic time-dependencies}, where energy eigenstates may not exist~\cite{PilatowskyCameo2023, Mark2024, PilatowskyCameo2024, Logaric2025}. 
As a simple example, the quantum state of a spin driven by randomly fluctuating fields can be expected to completely and uniformly explore the Hilbert space over time, i.e., achieve `complete Hilbert space ergodicity' (CHSE), but Ref.~\cite{PilatowskyCameo2023} showed nontrivially that a quantum system driven by a deterministic but aperiodic sequence of kicks based on the Fibonacci word also achieves CHSE, demonstrating that randomness in the drives is not essential.
Subsequent works~\cite{Mark2024, PilatowskyCameo2024} have then provided general conditions when CHSE, and more generally, when more nuanced levels of Hilbert space ergodicity (HSE) --- as defined by statistical indistinguishability of the distribution of quantum states over time to that of uniformly random states over the Hilbert space for {\it finite} moments --- may be achieved depending on the form of a given, fixed drive. Specifically, it was shown that ergodicity at deeper levels, i.e., statistical indistinguishability at higher moments, requires more independent frequencies underlying a drive, which can be understood as one measure of its `complexity'.  While a recent work~\cite{h6xy-zpx4} observed experimental signatures of HSE in the dynamics of a Rydberg atom quantum simulator (specifically, they studied the statistics of probabilities of computational bitstring measurement outcomes over time), a direct experimental observation of HSE and its different levels --- which necessitates probing the full temporal distribution of quantum states over the Hilbert space in time, and not just fluctuations along a particular measurement basis --- has to date not been achieved. 

In this work, we provide the first unambiguous experimental observation of a hierarchy of HSE achieved in the dynamics of a single quantum spin realized by a nitrogen-vacancy (NV) center in diamond, which results from driving it with different time-dependent modulations. 
% \lwq{I've left the revision this way. If acceptable, please delete the colored text.}
% OK, let's proceed with this. if we get complaints from Referees, we dial it back. 
These include a time-periodic (Floquet) drive, a time-quasiperiodic drive composed of kicks with smoothly modulated amplitudes, and the Fibonacci drive, a sequence of kicks based on the Fibonacci word. 
To realize these drives, we leveraged an isotope-purified diamond sample whose spin dephasing time achieves $68~\mathrm{\upmu s}$, and developed 99.99\% fidelity quantum logic gates to further suppress noise in quantum evolution, allowing us to apply nearly 1000 operations in total with high precision.
Furthermore, after each quantum operation, we perform full state tomography of the spin state, enabling us to follow each spin trajectory in detail and thus directly probe the distribution of time-evolved quantum states over long times.

We find evidence that the quantum spin does not exhibit HSE at any level when driven by the Floquet drive; while it exhibits HSE at the first moment but not higher for the smoothly kicked quasiperiodic drive; and lastly it exhibits behavior consistent with CHSE  for the Fibonacci drive: the statistical indistinguishability of the time-evolved quantum states to Haar random vectors is seen to continually decrease over all experimentally observable times and for all experimentally probable moments, regardless of initial state. 
To explain these findings, we theoretically analyze the nature of the drives and posit that they possess increasing levels of drive complexities, captured not only by the number of tones underlying them --- a notion elucidated by Ref.~\cite{PilatowskyCameo2024} --- but also other notions such as the regularity of the modulations of the kick amplitudes over time. 
We argue that increasing levels of complexity of the quantum drive will generically lead to more complicated trajectories through Hilbert space and hence increasing levels of HSE, in line with the experimental observations. 
Our work provides the first unambiguous experimental evidence for HSE, and unveils deeper, fine-grained forms with which closed quantum systems may reach equilibrium as well as the mechanisms underlying them, novel physics which goes beyond the standard framework of equilibration at the level of expectation values of observables.

%%%%% HSE and Drives %%%%%%%%%%%%%%%%%%%%%%%%%%%%%%
\section{Hilbert space ergodicity and drive protocols realized}

In this section, we recap the theoretical framework of Hilbert space ergodicity (HSE)~\cite{PilatowskyCameo2023, Mark2024, PilatowskyCameo2024, Logaric2025}, in particular explaining the measure used to probe the various levels of HSE achievable in the dynamics of a quantum system. We then introduce the three quantum drives studied in this work, which are used to drive our experimental system consisting of a single NV center in diamond.

\begin{figure*}
\centering
\includegraphics[width=1\columnwidth]{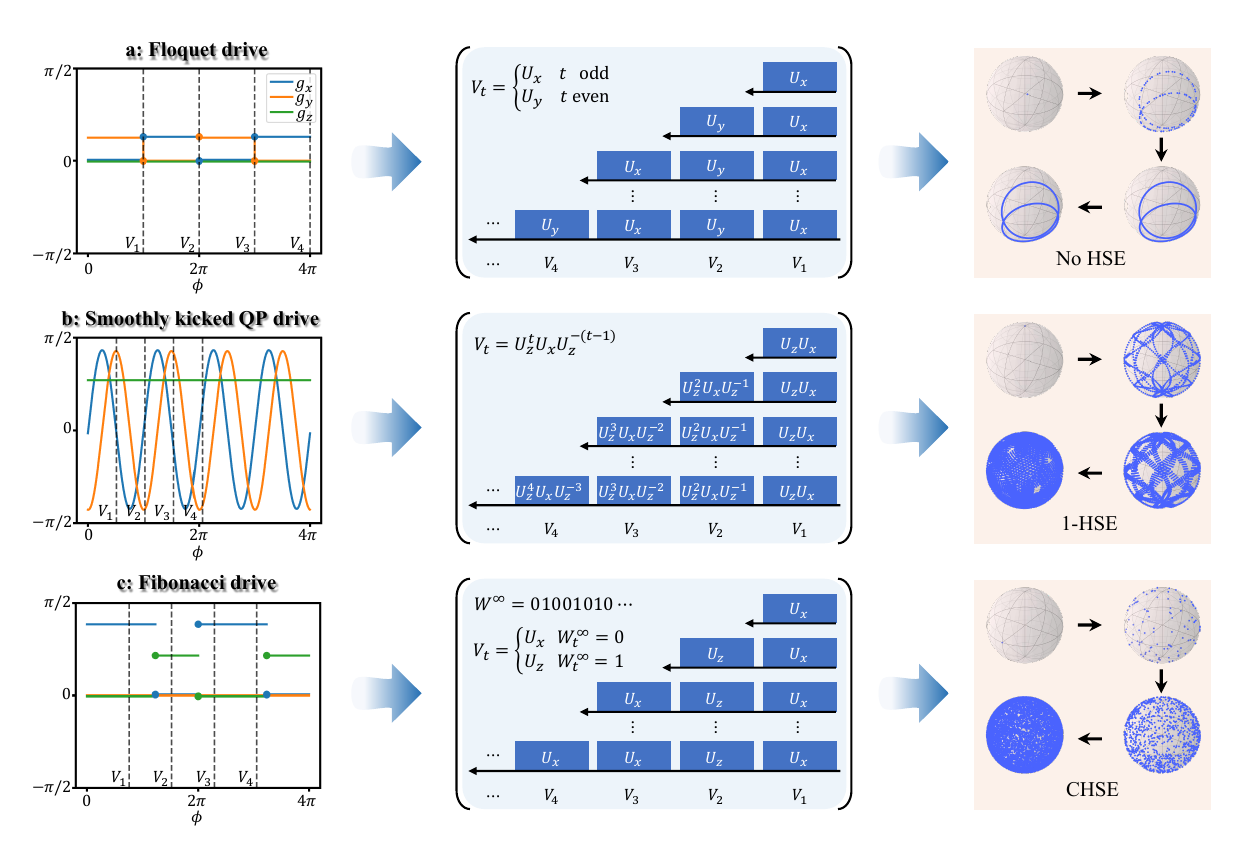}
\caption{
\textbf{Drive protocols realized and resulting dynamics over the Hilbert space.}
(a-c) show the Floquet drive, smoothly kicked QP drive, and Fibonacci drive protocols and resulting dynamics, respectively. The left panels depict the kick modulation vector $g(\phi)$. Upon evaluating them at times $\omega_2t$ (dashed vertical lines), this results in kicks $V_t$. The middle panels show the resulting unitary time-evolution operator $U(t)$ composed of the sequence of kicks (time runs from right to left). The right panels show numerically simulated examples of quantum dynamics on the Bloch sphere, in which we have chosen the initial states $|\theta,\varphi\rangle$ as $|\pi/3,\pi/5\rangle$ in (a), $|0,0\rangle$ in (b), and $|\pi/2,0\rangle$ in (c), where $(\theta,\varphi)$ are the standard polar and azimuthal angles defining a qubit state. One sees that the Floquet drive does not seem to explore the Hilbert space uniformly at all, while the smoothly kicked QP drive and Fibonacci drive do, though to differing degrees, as can already be seen by eye. These behaviors are quantified in the experimental results section and corroborated in the theoretical discussion section, as giving rise to ``no-HSE'', ``1-HSE'', and ``CHSE'' respectively. 
}
\label{Drive_protocols}	
\end{figure*}
Hilbert space ergodicity (HSE) concerns the statistical randomness of the sequence of time-evolved quantum states $\{ |\psi(t)\rangle \}_{t \geq 0}$  of a $d$-dimensional quantum system undergoing unitary dynamics by a drive Hamiltonian $H(t)$, called the `temporal ensemble'.
Namely, it asks if the system explores its ambient space, the Hilbert space, uniformly over time, probed by comparing the statistical similarity of the temporal ensemble to the ensemble of Haar random vectors, states in the Hilbert space whose distribution is invariant under any unitary rotation in SU($d$). Precisely, this similarity can be ascertained at various levels given by equality at different statistical moments $k \in \mathbb{N}$ of the ensembles, quantitatively captured by the vanishing of the trace distance
\begin{equation}
\Delta^{(k)}(T) := \frac{1}{2}\|\rho_T^{(k)}-\rho_{\rm{Haar}}^{(k)}\|_1
\label{kHSE}
\end{equation}
 at late observation times $T$. Above, $\| \cdot \|_1$ is the trace norm, while $\rho_T^{(k)}:=1/T\int_0^T dt  (|\psi(t)\rangle\langle\psi(t)|)^{\otimes k}$ is the $k$th moment of the finite-time temporal ensemble constructed from dynamics from an initial state $|\psi(0)\rangle$, and $\rho_{\rm{Haar}}^{(k)}:=\int d\phi (|\phi\rangle\langle\phi|)^{\otimes k}$ is the $k$th moment of the Haar ensemble with $d\phi$ the Haar measure on the unitary group SU($d$).

 If $\Delta^{(k)}(T)$ vanishes when dynamics is initialized from {\it any} state $|\psi(0)\rangle$, then this property is termed `$k$-HSE' (in practice, we can say this property holds whenever the trace distance drops below a fixed small but non-zero threshold $\epsilon > 0$). In quantum information parlance, this is the statement of the system's dynamics generating a quantum state $k$-design over time, agnostic to the initial state --- that is, that the sequence of quantum states in dynamics cannot be information-theoretically distinguished from random vectors even when given $k$ copies of each state. 
Note that $k$-HSE for $k\geq 2$ goes beyond the standard framework of equilibration at the level of time-averaged expectation values of an observable, captured only by 1-HSE. 
Indeed,  the various levels of HSE form a strict hierarchy: $k$-HSE implies $k'$-HSE whenever $k > k'$. In particular, when HSE holds for all $k$,  the system is said to exhibit {\it complete} Hilbert space ergodicity --- the time-evolved states cannot be statistically distinguished from Haar random ones through any information theoretic test. 
It is this hierarchical nature of ergodicities which underpins one of the motivations of this work: for which quantum dynamics can we expect deeper levels of HSE, and how do they depend on the form or properties of the driving Hamiltonian $H(t)$?

Here we experimentally study this question by driving a qubit (so $d= 2$) realized by a NV center in diamond, with three different time-dependent Hamiltonians that have distinct temporal profiles. 
In all cases, the driving Hamiltonian are of the form
\begin{align}
H(t) = \sum_{n \in \mathbb{N}}\delta(t-2\pi n/\omega_1) \vec{g}(\omega_2 t)\cdot \vec\sigma,
\label{eqn:Ht}
\end{align}
where we will set the first frequency component $\omega_1 = 2\pi$ for simplicity, $\vec\sigma$ is the vector of standard Pauli matrices, and $\vec{g}(\phi)$ a choice of Bloch vector defined on the circle $S^1 = [0,2\pi] \ni \phi$, which together with the second frequency component $\omega_2$ will determine the particular drive realized.
The Hamiltonian Eq.~\eqref{eqn:Ht} generates quantum dynamics in the form of a sequence of kicks $V_t =e^{-i \vec{g}(\omega_2 t) \cdot \vec\sigma}$ applied at   stroboscopic times $t \in \mathbb{N}$, namely the unitary time-evolution operator reads 
\begin{align}
U(t) = V_t \cdots V_2 V_1 =e^{-i \vec{g}(\omega_2 t) \cdot \vec\sigma} \cdots e^{-i \vec{g}(2\omega_2 ) \cdot \vec\sigma}e^{-i \vec{g}(\omega_2) \cdot \vec\sigma}.
\end{align}
One can understand these kicks as applying fields    $\vec{g}(\phi)$ evaluated at coordinates $\phi = \omega_2 t$ on the circle at integer time steps, see Fig.~\ref{Drive_protocols}; thus, $\vec{g}(\phi)$ parameterizes the way that the kicks are being modulated in time. 
Since quantum dynamics occurs at discrete times $t \in \mathbb{N}$, it is natural to modify the definition of the temporal ensemble used in the evaluation of $k$-HSE Eq.~\eqref{kHSE} to be those of the discrete set of states $\rho_T^{(k)}:= 1/T\sum_{t=0}^{T-1} (|\psi(t)\rangle \langle \psi(t)|)^{\otimes k}$) obtained in dynamics. 

The first drive we will realize is obtained by setting $\omega_2 = \pi$ and choosing the kick modulation vector
\begin{align}
\vec{g}(\phi) = \begin{cases}
\theta_y \hat{y} & 0 \leq \phi < \pi, \\
\theta_x \hat{x} & \pi \leq \phi < 2\pi.
\end{cases}
\label{eqn:Floquet}
\end{align}
This generates a time-evolution operator consisting of alternating kicks in the $x$ and $y$ directions: $U(t) = 
\underbrace{\cdots U_y U_x U_y U_x}_t$, 
where $U_\alpha = e^{-i\theta_\alpha \sigma_\alpha}$ ($\alpha = x,y,z$). Evidently, this describes a time-periodic drive, and hence we term it a `Floquet drive'. 
The second drive we will realize is defined by choosing $\omega_2 = (1+\sqrt{5})/2$ and kick modulation vector
\begin{align}
 \vec{g}({\phi}) = \frac{c}{\sin c}\begin{pmatrix}
\sin1 \cos(2\phi -  \omega_2)  \\
\sin1 \sin(2 \phi -\omega_2) \\
\cos 1 \sin(\omega_2)
\end{pmatrix},
\label{eqn:smooth}
\end{align}
with $\cos c = \cos \omega_2 \cos 1$. 
Note each kick $V_t = e^{-i \vec{g}(\omega_2 t)\cdot \sigma}$ can be shown to be equivalent to $V_{t}= U_z^t U_x U_z^{-(t-1)}$ with $\theta_x=1$, and $\theta_z=\omega_2$. 
% Each kick $V_t = e^{-i \vec{g}(\omega_2 t)\cdot \sigma}$ can be shown to be equivalent to $V_{t}= U_z U_{x^{(t-1)}}$ with $U_{x^{(t)}}:= e^{-i U_z^{t}\sigma_x U_z^{-t}}$ and $U_z := e^{-i \omega_2 \sigma_z}$. 
%\lwq{Wenwei, this definition is somewhat complex and not intuitive. Particularly, the placement of the U operator in the exponential function might initially suggest a typo error to readers, though careful verification would confirm its correctness. How about adopting the form shown in the figure? This would simplify the notation and enhance correspondence with the figure} 
%\textcolor{red}{it is possible but I personally find the version in the figure complex and convoluted..? In the current version here the idea that it is a  kick which rotates in the xy-direction followed by a kick in z. We can take a vote...(?) Note $U e^{i A} U^\dagger = e^{i UAU^\dagger}$ for any linear operator $A$. }
%\lwq{Since we only plotted $U_x$ and $U_z$ operations in the figure, no rotation in the XY-direction, so I will vote for $U_z^tU_xU_z^{-(t-1)}$. Zouwei said he also prefers this to avoid extra definitions, since we have defined rotation $U_\alpha$ previously.}
%\textcolor{red}{That's fine, let's go with the version presented in the figure then}
Since $\vec{g}(\phi)$  is a smooth function over the circle, and $\omega_2$ is incommensurate with $\omega_1$ resulting in the sequence of points $\{ \omega_2 t \}_{t=1,2,3,\cdots}$ densely filling the circle, the Hamiltonian can be understood to correspond to a `smoothly kicked quasiperiodic (QP)' drive (of two tones). 
The last drive we will realize is  defined by $\omega_2 = \pi(3-\sqrt{5})$ and the choice of  kick modulation vector
\begin{align}
\vec{g}(\phi) = 
\begin{cases}
\theta_x \hat{x} & 0 \leq \phi < 2\pi-\omega_2, \\
\theta_z \hat{z} & 2\pi-\omega_2 \leq \phi < 2\pi.
\end{cases}
\label{eqn:Fibonacci}
\end{align} 
The unitary time evolution operator at time $t$ reads $U(t) = \cdots U_x U_z U_x U_z U_x U_x U_z U_x$, which can also be understood as obtained from substitution of characters $0,1$ of the Fibonacci word $W^\infty = 01001010\cdots$ 
(generated by the concatenation rule on words $W^{(k+1)} = W^{(k)} W^{(k-1)}$ with $W^{(0)} = 1, W^{(1)} = 0$), with basic unitaries $U_x, U_z$ respectively. This Hamiltonian generates the `Fibonacci drive' of Ref.~\cite{PilatowskyCameo2023}, for which we expect CHSE. 

Fig.~\ref{Drive_protocols} shows an illustration of the definition of the three drives realized and the kick modulation vectors $\vec{g}(\phi)$ which underlie them, as described above. 
In addition, as a preliminary investigation, we show plots of the distributions that the temporal ensembles trace over the Hilbert space (here, the Bloch sphere) beginning from example initial states for the different drives, obtained from numerical simulations. 
One sees that the Floquet drive does not seem to explore the Hilbert space uniformly at all; while the smoothly kicked QP and Fibonacci drive do --- though to differing degrees, which can already be seen by eye. In the experimental results section, we will systematically quantify these differences as resulting in varying levels of HSE, namely ``no-HSE'', ``1-HSE'', and ``CHSE'', respectively, observations which will be further corroborated and explained with theoretical analyses.

%%%%% Experimental Setup %%%%%%%%%%%%%%%%%%%%%%%%%%%%%%
\section{Experimental Setup}

\begin{figure*}[t]
\centering
\includegraphics[width=1\columnwidth]{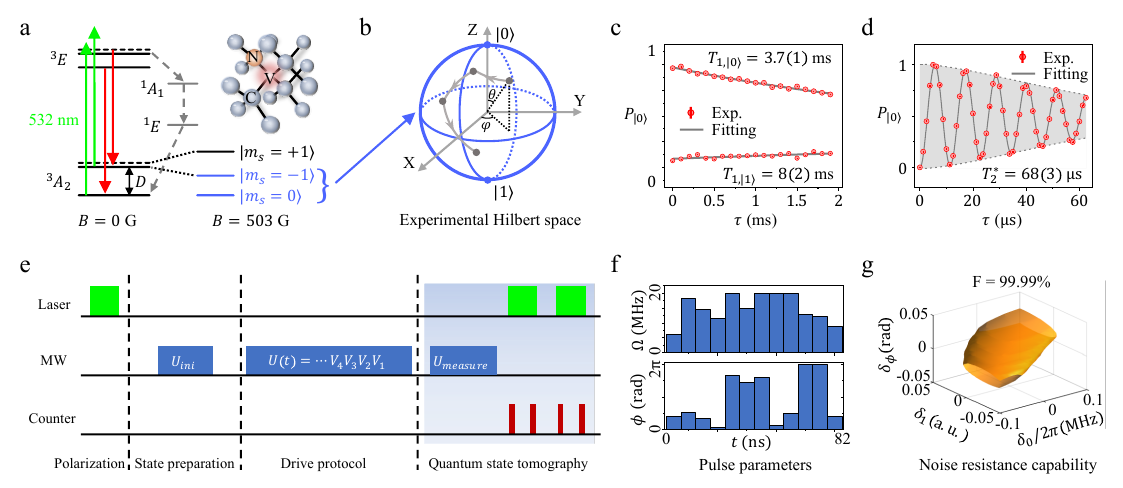}
\caption{
\textbf{The NV center system and the pulse scheme.}
(a) The lattice structure and energy levels of the NV center system. A magnetic field of 503 G, with its direction along the NV symmetry axis, was applied to split the $|m_s=\pm1\rangle$ states. (b) Experimental Hilbert space spanned by states $\ket{m_s=0,-1}$. The Hilbert-space ergodicity was explored by measuring the spin trajectory in the Bloch sphere. (c-d) Longitudinal and transverse relaxation times of the two-level system. $T_{1,|i\rangle}$ denotes the relaxation starting from state $|i\rangle$ with $i\in\{0,1\}$. The error bars are not visible because they fall within the data points. (e) Pulse scheme to measure the spin dynamics under different drive protocols. (f) Design of the high-fidelity quantum operations exemplified by $U_x=e^{-i 0.38\pi \sigma_x}$. The amplitudes and phases of the pulses are plotted. (g) Noise resistance capability of the designed operation against dephasing noise $\delta_0$, pulse amplitude fluctuation $\delta_1$, and pulse phase distortion noise $\delta_\phi$.}
\label{platform}	
\end{figure*}

Before presenting our findings, we describe our experimental setup. The quantum system we utilized to experimentally investigate Hilbert-space ergodicity consists of a single electron spin of a NV center in diamond, a point defect formed when a nitrogen atom replaces a carbon atom in the diamond lattice and pairs with an adjacent vacancy, as shown in Fig.~\ref{platform}a~\cite{Gali2008, Maze2011}. The vacancy contains six electrons and can be regarded as a spin-1 electron spin. The energy levels of the spin system are depicted in Fig.~\ref{platform}a. The ground state ${}^{3}A_2$ is a spin triple state with energy levels denoted by $|m_s = 0,\pm1\rangle$. These levels are separated by an internal zero-field splitting of $D = 2.87$ GHz, in addition to the Zeeman splitting caused by an external magnetic field $B$ set at 503 G. Initialization and readout of the spin can be realized through a spin-state-dependent intersystem crossing process, which involves the first excited ${}^{3}E$ and two singlet states ${}^{1}A_1$ and ${}^{1}E$, by applying a 532 nm laser pulse~\cite{Jacques2009, Thiering2018}.

Spin states $|m_s = 0, -1\rangle$ were chosen to form a two-level system (TLS) and span the experimental Hilbert space as shown in Fig.~\ref{platform}b. We relabel them as $|0\rangle$ and $|1\rangle$ hereinafter for simplicity. The diamond sample was synthesized using the chemical vapor deposition method~\cite{Achard2020, Luo2022}. During the synthesis, we employed purified raw materials, reducing the proportion of ${}^{13}\rm{C}$ isotope from its natural abundance of 1.1\% to less than 0.1\%. This reduces the number of nuclear spins nearby the NV center, therefore mitigating noise sources affecting the TLS. The longitudinal and transverse relaxation times of the TLS were measured and plotted in Fig.~\ref{platform}c and d. The fitting results show that they are $T_{1,|0\rangle} = 3.7(1)$ ms, $T_{1,|1\rangle} = 8(2)$ ms, and $T_2^* = 68(3) \upmu$s, respectively. 

The pulse scheme to realize the different drive protocols and measure the corresponding spin dynamics is depicted in Fig.~\ref{platform}e. We begin by polarizing the spin to $|0\rangle$ via a 532 nm laser pulse. Then, operation $U_{ini}$ was executed to prepare the spin to different initial states $|\psi(0)\rangle=U_{ini}|0\rangle$.
Afterward, the different target drive protocols were applied to dynamically evolve the spin. 
At a desired time-step in time-evolution, different operations $U_{measure}$ followed by laser and counter pulses were performed to realize quantum state tomography of the spin state. 

We note that experimentally, imperfections in quantum manipulations can cause the spin dynamics to deviate from the idealized trajectories and fail to reveal the ergodic property of the drive. 
To address this, we have developed high-fidelity quantum logic gates for the involved $U_{x/y/z}$ operations based on quantum optimal control methods~\cite{Rembold2020}. It is worth mentioning that in addition to the dephasing and pulse amplitude noises, we also took into account pulse phase noise, which was frequently overlooked in previous studies~\cite{Rong2015, Xie2023}. In Fig.~\ref{platform}f we present an example operation $U_x=e^{-i0.38\pi \sigma_x}$ implemented which has a duration of 82 ns and consists of 12 pulses with changing amplitudes and phases. The performance of this operation is depicted in Fig.~\ref{platform}g, which shows that within a large parameter space, the fidelity exceeds 99.99\%. The experiment was carried out in a NV center based optically detected magnetic resonance setup~\cite{Doherty2012, Wrachtrup2016, Barry2020, Wolfowicz2021, Du2024}. The precise configuration of this setup and other experimental details are presented in the Methods section.

\section{Experimental results}

\begin{figure*}[!htb]
\centering
\includegraphics[width=1\columnwidth]{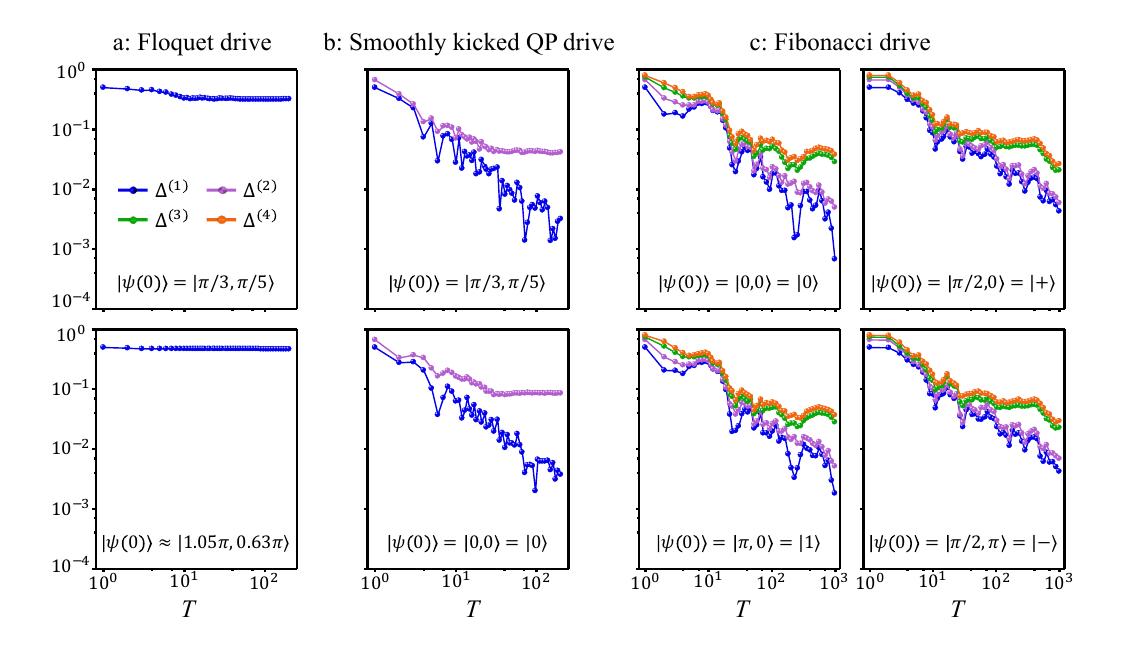}
\caption{ 
\textbf{Experimental results.} (a) Experimentally determined trace distance $\Delta^{(1)}$ of the Floquet drive beginning from initial states $|\psi(0)\rangle$ as indicated at the bottom of the plots; $|\theta,\varphi\rangle$ correspond to a Bloch state with polar and azimuthal angles $\theta,\varphi$ respectively. Initial state in the bottom panel corresponds to an eigenstate of the Floquet unitary. In both top and bottom panels, $\Delta^{(1)}$ plateaus quickly, indicating that the dynamics does not exhibit HSE for any $k$. (b) Trace distances $\Delta^{(1)}$ and $\Delta^{(2)}$ plot of the smoothly kicked  QP drive for various initial states.  $\Delta^{(1)}$ is seen to continually decay even up to the largest observation time, independent of initial state, while $\Delta^{(2)}$ is seen to plateau after an initial decay. This suggests the smoothly kicked QP drive achieves a limited form of ergodicity, namely 1-HSE, but not 2-HSE or higher. (c) Trace distance $\Delta^{(k)}$ plots of the Fibonacci drive for $k = 1,2,3,4$. The trace distance of all moments measured is seen to decay throughout the observation window, indicating signatures of {\it complete} HSE (CHSE). 
}
\label{Exp_Results}	
\end{figure*}

We now present our experimental findings. 
For the Floquet drive, 
we take rotation angles $\theta_x = \theta_y = \pi/8$ and total evolution time to be 200 time steps.
The top panel of Fig.~\ref{Exp_Results}a shows the experimentally measured trace distance $\Delta^{(1)}(T)$ of the first moment beginning from a randomly chosen initial state (Bloch angles indicated in the plot): we see there is an initial decay, but soon it   plateaus to a non-zero value $\approx 0.32$.
This directly quantifies the non-ergodicity seen in the trace of the quantum states over the Bloch sphere in Fig.~\ref{Drive_protocols}, and indicates that our Floquet drive does not even exhibit  1-HSE, let alone $k$-HSE. 
Intuitively, the origin of the non-ergodicity can be traced to the presence of quasienergy eigenstates of the Floquet operator $U_F  = U_y U_x$, which forms the basic building blocks of the total time-evolution operator $U(t)$, and which are stationary states after every Floquet cycle (here after every two time integer steps).
To confirm this, we repeat the same experiment beginning from an initial state which is one of the quasienergy states of $U_F$ (numerically determined); results are shown in Fig.~\ref{Exp_Results}a, bottom panel. As expected, the trace distance for the first moment decays even less than for the previous initial state,   plateauing at a higher value of $\approx0.47$.

The results for the smoothly kicked QP drive  are depicted in Fig.~\ref{Exp_Results}b, with different initial states indicated in the top and bottom panels. Within the observation time interval, we find that the trace distance for the first moment $\Delta^{(1)}$ 
exhibits a continual power-law decay for all times measured, and also for all initial states  (this is also true for  other states investigated; not shown).
In contrast, the measured $\Delta^{(2)}$ is seen to decrease during the early stages of time evolution for both initial states, however it eventually stabilizes to a non-zero value. These findings strongly suggest that the smoothly kicked QP drive exhibits 1-HSE, but not 2-HSE or higher. 
In the Methods section, we rigorously show that indeed, the smoothly kicked QP  drive, defined by  Eq.~\eqref{eqn:smooth},  forms a  quantum state 1-design in dynamics, provided that $\omega_2$ is incommensurate with respect to both $2\pi$ and $1$ (we note these conditions are met in our experiments).  
The sketch of the proof relies on noting that the the dynamics can be be re-written as $U(t) = e^{-i \omega_2t\sigma_z} e^{-i t \sigma_x}$, and considering the effect of the time-averaging channel on a given density matrix: $\mathcal{T}_T[\rho]:= 1/T \sum_{t=0}^{T-1}U(t)\rho U^\dagger(t)$. 
By the incommensurability assumptions on $\omega_2$, we can evaluate the action of the time-averaging channel, finding that it is equal to the consecutive application of a maximally dephasing channel in the $x$ then $z$ directions, which results in a maximally mixed state {\it regardless} of initial state. Details are provided in the Methods section. 
 
Finally, the experimental results for the Fibonacci drive  are illustrated in Fig.~\ref{Exp_Results}c for a choice of rotation parameters  $\theta_x = 0.38\pi$ and  $\theta_z = 0.22\pi$. Spin dynamics starting from four different initial states, $|\psi(0)\rangle\ = |0\rangle, |1\rangle, |+\rangle$ and $|-\rangle$, were studied. We applied up to 987 operations (corresponding to the 15th-order Fibonacci sequence) and experimentally calculated the trace distances for moments up to the fourth order. 
The subplots in Fig.~\ref{Exp_Results}c show that despite the presence of some oscillations in the obtained $\Delta^{(k)}$, the overall trend is one of persistent decay for all $k$, consistent with a power-law. These observations are indicative of the emergence of CHSE,  which had been theoretically predicted by Ref.~\cite{PilatowskyCameo2023} for   Fibonacci drives with generic drive parameters. 

\begin{figure}[!h]
\centering
\includegraphics[width=1\columnwidth]{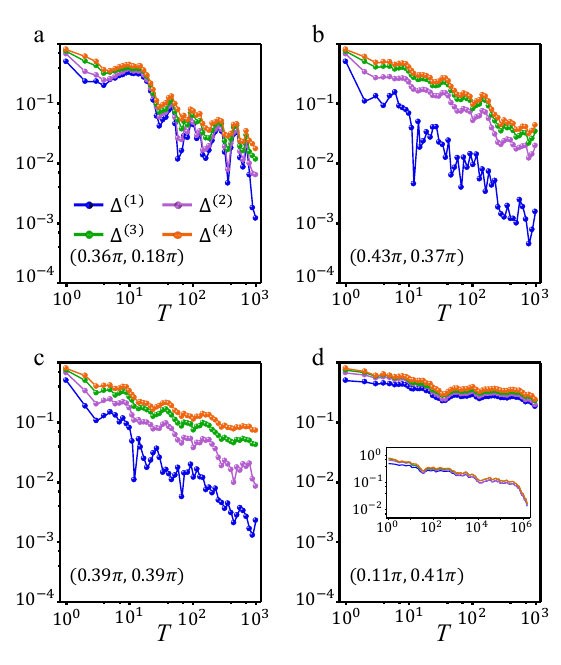}
\caption{ 
\textbf{Experimental results for Fibonacci drives with other system parameters.} (a-d) The trace distances of Fibonacci drive with other parameters (listed in each plot) were measured. $\Delta^{(k)}$ for all measured $k$ is seen to decay for all observation times, indicating CHSE. Inset in (d) shows independent numerical simulations with the same system parameters for longer times, demonstrating that $\Delta^{(k)}$ indeed decays continually with no evident saturation. }
\label{More_Results}		
\end{figure}
 
To further investigate that the observed CHSE  is universal and robust to system parameters of the Fibonacci drive, we re-ran the experiments with different drive parameters $(\theta_x, \theta_z)=(0.43\pi, 0.37\pi), (0.36\pi, 0.18\pi), (0.39\pi, 0.39\pi)$, and $(0.11\pi, 0.41\pi)$. Here the initial state was fixed to $|\psi(0)\rangle=|0\rangle$ in all cases. The obtained curves for $\Delta^{(k)}$ for $k=1,2,3,4$ are presented in Fig.~\ref{More_Results}. The data shows that all measured trace distances continually decay as observation time $T$ increases with no evidence of saturation, though the rate of decay vary greatly. 
In particular,  the traces distances in Fig.~\ref{More_Results}d for parameters $(0.11\pi,0.41\pi)$ are seen to decrease exceptionally slowly. Nevertheless, independent numerical simulations (shown in the inset) run for times longer than the experimentally accessible observation times shows that there is still a sustained decay regardless. Together, these observations support compellingly the completely Hilbert-space ergodic nature of the Fibonacci drive.

%%%%% Discussion %%%%%%%%%%%%%%%%%%%%%%%%%%%%%%
\section{Discussion}
We now turn to a theoretical discussion of the origin and implications of our experimental findings, namely that increasing levels of HSE are seen for the Floquet, smoothly-kicked QP, and Fibonacci drives, respectively. While we have a concrete understanding of the levels of HSE these specific drives realize (namely, the presence of quasienergy states resulting in no HSE for the Floquet drive, and mathematical proofs showing 1-HSE and CHSE for the latter two), we now ask: 
can these levels of Hilbert space ergodicity  already be anticipated based on certain general properties of these drives, principles which can then be used to understand the ergodic behavior of other driving sequences? 

As mentioned in the introduction, Ref.~\cite{PilatowskyCameo2024} has provided such a theoretical analysis, at least along one front: there, it was argued that the number of fundamental frequencies underlying a given drive  constitute a notion of its `complexity' and determine its ergodic properties. Intuitively, more frequencies   result in   trajectories of a quantum state  through Hilbert space that are more complicated and can hence cover it more uniformly; conversely, fewer frequencies limit the ability of path of a quantum state to   explore the space. Indeed, this is rigorized by no-go theorems of Ref.~\cite{PilatowskyCameo2024} which forbid CHSE in a given $d$-level quantum system if driven by too few tones, under the assumption of existence of quasienergy states. 
In our present case, the Floquet drive, being time-periodic, consists of a single underlying frequency (this can mathematically be traced to the fact that the second frequency component $\omega_2$ is commensurate with respect to the first frequency component $\omega_1$), while the smoothly-kicked QP drive and Fibonacci drives are both two-frequency time-quasiperiodic drives (since $\omega_2$ and $\omega_1$ are irrationally related). Thus, the latter two drives' can be said to be more complex (in the number of tones sense) than the first and should be expected to achieve a higher level of ergodicity. This is indeed what we see experimentally.

However, both the smoothly-kicked QP drive and Fibonacci drive consist of a similar number of underlying frequencies (they are both two-tone QP drives), and hence are equally complex from this point of view. What then is the distinction between them? 
Analyzing the form of the kick modulation vectors $\vec{g}(\phi)$ in   Eq.~\eqref{eqn:smooth} and Eq.~\eqref{eqn:Fibonacci} defining them, we observe that a key difference is that $\vec{g}(\phi)$ for the former is a smooth function over the circle, resulting in a sequences of kicks whose amplitudes are being modulated smoothly over time (and hence giving rise to the drive's name); while $\vec{g}(\phi)$ for the latter is not smooth, with step discontinuities,  see Fig.~\ref{Drive_protocols}. This results in kicks whose amplitudes are being modulated discontinuously. 
While these differences seem technical, we argue that they can have physical consequences --- the regularity of modulations over time determine the drives' power spectra (i.e., distribution of frequencies): smoothly modulated drives have more localized spectra while discontinuously modulated drives have more delocalized spectra.  These determine the energy that a quantum system can absorb or emit from the drive that is used to transition between different quantum states, which in turn affects its ability to explore the Hilbert space. We should thus expect that
% the smooth or discontinuous nature of the kick modulations   is another factor determining the degree of quantum ergodicity achievable, with 
more(less) regular kicks result in less(more) degrees of HSE, which is indeed what we see in experiments. 

Therefore, our investigations have revealed that regularity of a drive's modulation in time constitutes another form of a drive's `complexity', which affect the level of ergodicity realizable. We note our discussion is in line with past works in the quantum chaos literature showing how differences in the regularity of drive modulations can lead to fundamentally distinct classes of responses of observables, namely whether the spectra of their correlation functions is discrete or continuous in nature~\cite{PhysRevE.51.1762, Arkady}.

%%%%% Conclusion and outlook %%%%%%%%%%%%%%%%%%%%%%%%
\section{Conclusion and outlook}
In this work, we have provided first unambiguous experimental observation of Hilbert-space ergodicity, a novel form of quantum ergodicity rooted in statistical pseudorandomness of quantum states generated in dynamics. 
Such a study was made possible by leveraging a single electron spin realized by a NV center in a diamond sample which is isotopically purified, which together with high-fidelity quantum operations developed and continuous tomography of the quantum system, allow for the collection of statistics of time-evolved quantum states over long times and hence probe their ergodicity over the Hilbert space. 
We implemented three drive protocols, and found that the quantum spin achieves hierarchical levels of HSE in the ensuing dynamics, ranging from no HSE, to partial HSE, to complete HSE. We theoretically attribute these increasing levels of ergodicity to increasing levels of complexity of the implemented drives, which we argue come in the form of the number of underlying frequencies, as well as the regularity of the modulations of kicks over time. 

Our work illustrates that closed quantum systems can achieve equilibrium in deeper, more fine-grained ways that go beyond the standard framework of time-averaged observables, and further through nuanced physical mechanisms that depend on the nature of the driving Hamiltonian, paving the ground for further investigations.
For example, formalizing how the different notions of drive complexities uncovered in this work and beyond, result in quantitative statements on the different degrees of HSE achievable, would be highly desirable. It would also be interesting to extend the study to systems with symmetries like energy conservation, and bridge the gap between dynamical notions of ergodicity like in  HSE, and the more conventional, static notions as defined by statistical pseudorandomness of stationary states,  in order to achieve a unified framework of quantum ergodicity. 

\begin{acknowledgments}
This work was supported by the National Natural Science Foundation of China (Grants No. 12475027, 124B1037, T2388102, 12261160569), the Fundamental Research Funds for the Central Universities (Grant No. 226-2024-00011), Technology Department of Zhejiang Province (2025C02027), and Zhejiang Key Laboratory of R$\&$D and Application of Cutting-edge Scientific Instruments.
W.~W.~H.~is supported by the National Research Foundation (NRF), Singapore, through the NRF Felllowship NRF-NRFF15-2023-0008, and through the National Quantum Office, hosted in A*STAR, under its Centre for Quantum Technologies Funding Initiative (S24Q2d0009). The authors thank Liang Zhang for insightful discussions.
\end{acknowledgments}

\appendix

\section{Experiment setup} 
\begin{figure}[htbp]
\centering
\includegraphics[width=1\columnwidth]{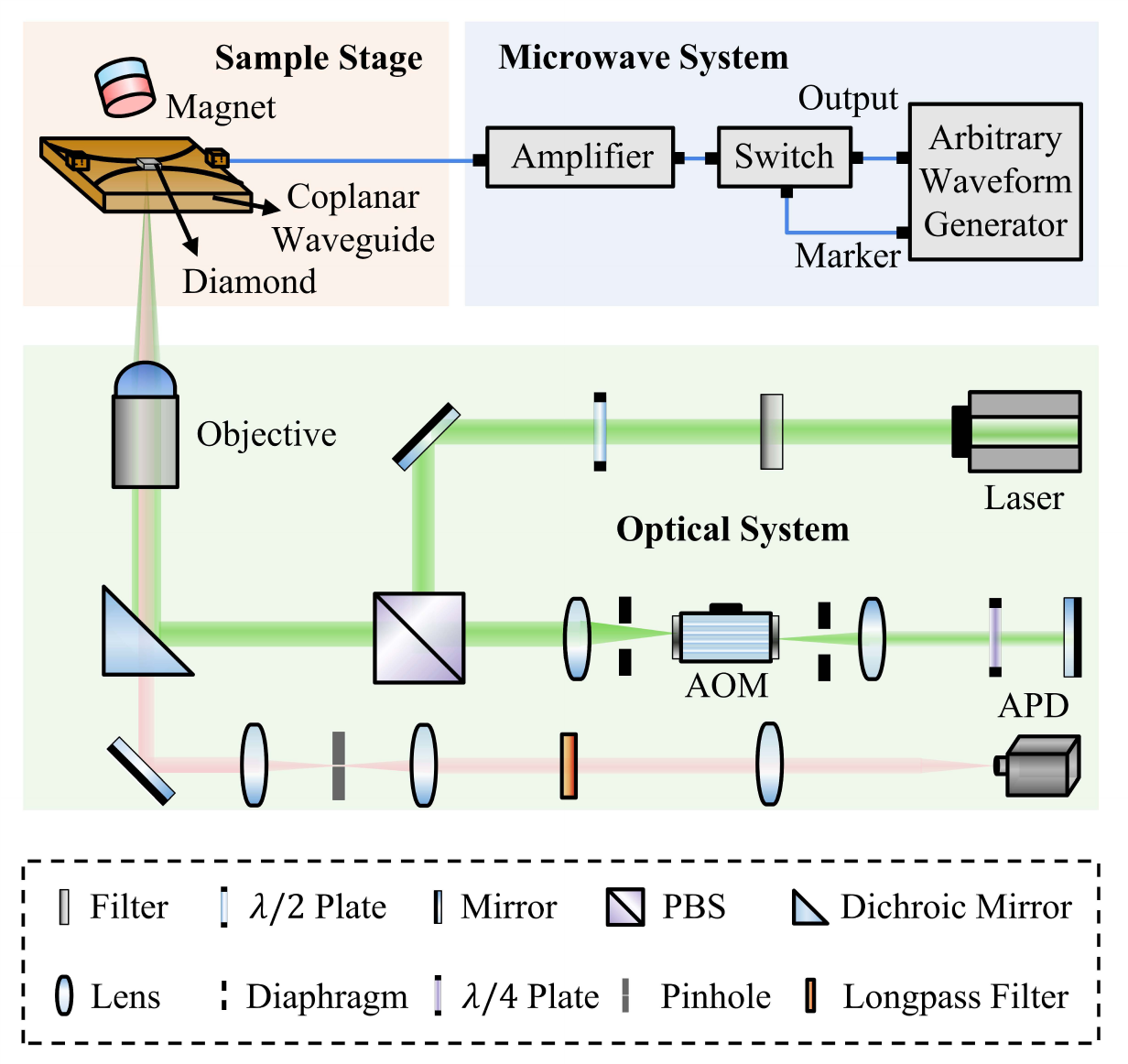}
\caption{ 
\textbf{The NV center based optically detected magnetic resonance setup.} The diamond sample was placed in a homemade coplanar waveguide. The optical system was utilized to pump and readout the spin. The microwave system outputs pulses to manipulate the spin state. }
\label{Setup}	
\end{figure}

The experiment was carried out in an optically detected magnetic resonance setup. The setup comprises three parts: the sample stage, the optical system, and the microwave system, as shown in Extended Data Fig. \ref{Setup}. The sample stage provides support for the diamond and incorporates a magnetic device, where the strength and direction of the magnetic field are electronically controllable (Lbtek EM-LSS65-30C1). The optical system was used to initialize and readout the spin state. Specifically, a 532 nm laser (CNI, MGL-III-532nm) emits a continuous beam modulated by an acousto-optic modulator (Gooch\&Housego, AUT-AOM 3200-1214), then focused by an objective (Olympus, LMPLFLN50X) to pump the NV center. Photoluminescence collected by the same objective passes through a pinhole (Thorlabs, P30K) and long-pass filter (Thorlabs, FELH0650) before detection by a single-photon detector (Excelitas, SPCM-AQRH-44-BR1). The microwave system generates pulses to manipulate spin states, enabling the implementation of different drive protocols. Microwave pulses from an arbitrary waveform generator (Keysight, 8190A) are routed through a switch (Mini circuit, ZASWA-2-50DRA+), amplified (Mini circuit, ZHL-15W-422-S+), and delivered via a homemade coplanar waveguide to manipulate the spin state.

\section{Measurement of the spin trajectory}
\begin{figure}[htbp]
\centering
\includegraphics[width=1\columnwidth]{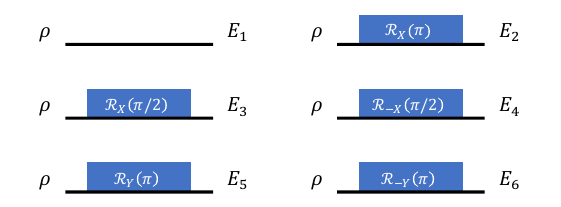}
\caption{ 
\textbf{Pulse sequences for the quantum state tomography.} Six pulse sequences were performed to measure different elements of the density matrix. Here, $\mathcal{R}_{m}(n)$ denotes spin rotation around axis $m$ by angle $n$.}
\label{tomography}	
\end{figure}

Experimentally, the density matrix after a certain number of quantum operations was measured by exploiting differential photoluminescence (PL) rates between distinct spin states. Let $l_0$ ($l_1$) represent the PL rate corresponding to the spin state $|0\rangle$ ($|1\rangle$), the expected PL rate for density matrix
\begin{equation}
    \rho = \begin{bmatrix}
        p_0 & \alpha + \beta i\\
        \alpha - \beta i & p_1
    \end{bmatrix}
    ,\quad p_0+p_1 = 1
\end{equation}
 is given by $E_1 = p_0l_0+p_1l_1$. Implementing the pulse sequences provided in Extended Data Fig. \ref{tomography}, yields the following equations:
 \begin{equation}
    \left\{
        \begin{aligned}
            &l_0p_0 + l_1p_1 = E_1\\
            &l_0p_1 + l_1p_0 = E_2\\
            &l_0\left(\frac{p_0+p_1-2\beta}{2}\right) + l_1\left(\frac{p_0+p_1+2\beta}{2}\right) = E_3\\
            &l_0\left(\frac{p_0+p_1+2\beta}{2}\right) + l_1\left(\frac{p_0+p_1-2\beta}{2}\right) = E_4\\  
            &l_0\left(\frac{p_0+p_1-2\alpha}{2}\right) + l_1\left(\frac{p_0+p_1+2\alpha}{2}\right) = E_5\\ 
            &l_0\left(\frac{p_0+p_1+2\alpha}{2}\right) + l_1\left(\frac{p_0+p_1-2\alpha}{2}\right) = E_6\\ 
        \end{aligned}
    \right.
\end{equation}
Therefore, the density matrix $\rho$ can be solved as:
\begin{equation}
    \left\{
        \begin{aligned}
            &p_0 = \frac{1}{2} + \frac{E_1-E_2}{2L_{01}}\\
            &p_1 = 1-p_0\\
            &\alpha   = \frac{E_3-E_4}{2L_{01}}\\
            &\beta   = \frac{E_5-E_6}{2L_{01}}
        \end{aligned}
    \right.
\end{equation}
where $L_{01}=l_0-l_1$ and can be obtained via Rabi oscillation experiment.

It should be noted that the electron spin cannot be fully polarized experimentally. Typically, the polarization efficiency $p_e$ is around 0.92, resulting in the measured spin states being mixed. The relationship between these mixed states and their corresponding pure states is given by 
\begin{equation}
    \rho_{\rm{mixed}} = \rho_{\rm{pure}} - (1-p_e)\cdot I_2  ,
    \label{mixed_state}
\end{equation}
where the identity component typically suppresses the spin dynamics from exhibiting HSE of the second order or higher. Therefore, based on Eq. \ref{mixed_state}, we reconstructed the spin trajectories on the Bloch sphere using experimental data to investigate their Hilbert-space ergodicity.
\\

\section{The smoothly kicked quasiperiodic drive exhibits 1-HSE.}
In this section, we rigorously prove that the smoothly kicked quasiperiodic drive realized in the paper exhibits 1-HSE, under the conditions that $\omega_2$ is incommensurate with respect to both $2\pi$ and $1$. 

It can be easily shown from the definition of the drive
% $U(t) = (U_z U_{x^{(t-1)}})\cdots (U_z U_{x^{(1)}} )(U_z U_{x^{(0)}})$ where $U_{x^{(t)}}:= e^{-i U_z^{t}\sigma_x U_z^{-t}}$ 
that equivalently $U(t) = e^{-i \omega_2 t \sigma_z} e^{-i t \sigma_x}$. 
Consider the 1-twirl defined by it in time: for any operator $O$, the 1-twirl channel $\mathcal{T}_T$ up to finite time $T$ is defined as
\begin{align}
\mathcal{T}_T[O] := \frac{1}{T}\sum_{t=0}^{T-1} U(t) O U^\dagger(t),
\end{align}
which is conventionally also called the `time-averaging channel'. 
Noting that $U(t)$ can itself be understood as obtained from evaluating a parent unitary on the two-torus
\begin{align}
U(t) & = V(t,\omega_2t), \\
V(\phi_1,\phi_2) & := e^{-i \phi_2 \sigma_z} e^{-i \phi_1 \sigma_x}, 
\end{align}
and that $\{ (1,\omega_2) t\}_{t \in \mathbb{N}}$ define a sequence of points which ergodically fills the torus with respect to the uniform measure, owing to the assumption of incommensurability of $\omega_2$ with respect to $1$ and $2\pi$, we can evaluate the infinite time 1-twirl $\mathcal{T} := \lim_{T \to \infty} \mathcal{T}_T$ as
\begin{align}
    \mathcal{T}[O] &=  \iint \frac{d\phi_2 d\phi_1}{(2\pi)^2} V(\phi_1,\phi_2)OV^\dagger(\phi_1,\phi_2) \\
    &= \int \frac{d\phi_2}{2\pi} e^{-i \phi_2 \sigma_z}
    \left( \int \frac{d\phi_1}{2\pi} e^{-i \phi_1 \sigma_x} O e^{i \phi_1 \sigma_x}  \right) e^{i \phi_2 \sigma_z}.
\end{align}
We recognize this to be the successive application of two quantum channels,
\begin{align}
    \mathcal{T}[O] = \mathcal{N}_z \circ \mathcal{N}_x [ O ],
\end{align}
with $\mathcal{N}_x$ and $\mathcal{N}_z$ defined in the obvious fashion, which is a concatenation of the symmetric bit-flip $\mathcal{N}_x$ channel and symmetric phase-flip $\mathcal{N}_z$ channels.

Taking $O$ to be some density matrix $\rho$, the bit-flip channel projects the Bloch vector of $\rho$ onto the $x$-axis, while the phase-flip channel further projects it onto the $z$-axis. Since the two projections are orthogonal, the end result is that regardless of initial Bloch vector, the resulting Bloch vector has zero length --- the combined channel is a maximally depolarizing channel. That is,
\begin{align}
    \mathcal{T}[\rho] = \int_{U \sim \text{Haar}(\text{SU}(2))} dU U \rho U^\dagger =  \frac{\mathbb{I}}{2}
\end{align}
for any   density matrix $\rho$.
This proves that the dynamics by the smooth quasiperiodic drive is 1-HSE.  
Note that the one can easily show that the smoothly kicked QP drive is not 2-HSE by finding an initial state which is not uniformly ergodic over the Hilbert space at the second moment.

\bibliography{CHSE_NV}

\end{document}